\documentclass{article}

\usepackage{arxiv}

\usepackage[utf8x]{inputenc} 
\usepackage[T1]{fontenc}    
\usepackage{hyperref}       
\usepackage{url}            
\usepackage{booktabs}       
\usepackage{amsfonts}       
\usepackage{nicefrac}       
\usepackage{microtype}      
\usepackage{lipsum}
\usepackage{xcolor}
\usepackage{graphicx}
\usepackage{setspace} 
\usepackage{amsmath}
\usepackage{array}
\usepackage{mdwmath}
\usepackage{mdwtab}
\usepackage{eqparbox}
\usepackage{url}
\usepackage{multirow}
\usepackage{textcomp}
\usepackage{diagbox}

\title{Adverse event enrichment tests using VAERS}

\author{
  Shuoran Li\\
  Department of Statistics\\
  University of Michigan\\
  Ann Arbor, MI 48109 \\
  \texttt{shuoranl@umich.edu} \\
  \And 
  Lili Zhao\\
  Department of Biostatistics\\
  University of Michigan\\
  Ann Arbor, MI 48109 \\
  \texttt{zhaolili@umich.edu} \\
}

\begin{document}
\maketitle

\begin{abstract}
Vaccination safety is critical for individual and public health. Many existing methods have been  used to conduct safety studies with the VAERS (Vaccine Adverse Event Reporting System) database. However, these methods frequently identify many adverse event (AE) signals and they are often hard to interpret in a biological context. The AE ontology introduces biologically  meaningful  structures  to  the  VAERS  database  by  connecting  similar AEs, which provides meaningful interpretation for the underlying safety issues.  In  this  paper,  we  develop  rigorous  statistical  methods to identify ``interesting" AE groups by  performing AE enrichment analysis. We extend existing gene enrichment tests to perform AE enrichment analysis. Unlike the continuous gene expression data, AE data are counts. Therefore, AE data has many zeros and ties. We propose two enrichment tests, AEFisher and AEKS. AEFisher is a modified Fisher's exact test based on pre-selected significant AEs, while AEKS is based on a modified Kolmogorov–Smirnov statistic. Both tests incorporate the special features of the AE data. The proposed methods were evaluated using simulation studies and were further illustrated on two studies using VAERS data. By appropriately addressing the issues of ties and excessive zeros in AE count data, our enrichment tests performed well as demonstrated by simulation studies and analyses of VAERS data. The proposed methods were implemented in R package AEenrich and can be installed from the Comprehensive R Archive Network, CRAN.
\end{abstract}


\section{Background}

The Centers for Disease Control and Prevention (CDC) and the U.S. Food and Drug Administration (FDA) conduct post-licensure vaccine safety monitoring using the Vaccine Adverse Event Reporting System (VAERS) \cite{Varricchio:2004,Shimabukuro:2015}. VAERS accepts spontaneous reports of suspected vaccine adverse events after administration of any vaccine licensed in the United States from 1990 to present. As a national public health surveillance resource, VAERS is a key component in ensuring the safety of vaccines. 

Numerous methods have been used to conduct safety studies with the VAERS database \cite{DuMouchel:1999,Evans:2001,van:2002,Bate:1998,Orre:2000,Noren:2006,DuMouchel:2001,Szarfman:2002,Kulldorff:2011,Davis:2005,Li:2009,Li:2014,Kulldorff:2013}. In these methods, a contingency table is generally created to display counts for all vaccine and adverse event pairs during a specified time period. In this table, each row represents a vaccine and each column represents an adverse event (AE). Each cell in the table contains the number of VAERS reports that mention both that vaccine and that event for a defined period. A statistical measure is then calculated to quantify the association between an adverse event and a vaccine. A large value of the measure shows a strong association, which might indicate a vaccine safety problem (called ``signal"). A signal is considered evidence that an adverse event might be caused by vaccination and warrants further investigation or action. However, these methods frequently identify many AE signals and they are often hard to interpret in a biological context.

Adverse events are naturally related; for example, events of retching, dysphagia and reflux are all related to an abnormal digestive system. The AE ontology introduce biologically meaningful structures to the VAERS database by connecting similar AEs, which provides meaningful interpretation for the underlying safety issues. The largest resource for describing AE relationships is MedDRA (Medical Dictionary for Regulatory Activities) \cite{Mozzicato:2009}.  It has a five level hierarchy.  VAERS uses the second lowest term, ``Preferred Terms" (PT), which is a distinct descriptor for a symptom, sign and disease. Related PTs are grouped into higher-level AE terms, including ``High Level Group Terms" (HLGT) and ``System Organ Classes" (SOC).  Higher layers of HLGT and SOC represent biologically and clinically meaningful categories for the AEs observed on the lower PT level. The AE ontology has been used to classify AE signals \cite{Zhang:2013,Marcos:2013,Sarntivijai:2012,Guo:2016}. For example, \cite{Zhang:2013} showed that most AE signals identified on the PT level were found to be in behavior/neurological AEs on the SOC level. However, these findings are based on an \textit{ad-hoc} strategy of comparing proportions of signaled AEs between AE groups. In this paper, we present rigorous statistical methods to identify groups of AEs that are associated with a vaccine of interest and quantify AE group uncertainty in the enrichment analysis.

Over the last few decades, bioinformatics methods have used gene ontology to systematically dissect large gene lists in order to assemble a summary of the most enriched and pertinent biology. The basic idea in the most traditional strategy for gene enrichment analysis is to take the user’s pre-selected  significant genes, and then compare difference between the proportion of significant genes that fall into the gene set and the proportion of significant genes that do not fall into the gene set \cite{Zeeberg:2003}. A more recent approach is the Gene Set Enrichment Analysis (GSEA) method \cite{Subramanian:2005, Mootha:2003}, which is using gene ranks based on a difference measure, such as fold change, rather than a ``cut-off" strategy based on gene significance. In GSEA, the distribution of gene ranks from the gene set is compared against the distribution for the rest of the genes by using the enrichment score (ES) based on a Kolmogorov–Smirnov statistic.

However, there are two important issues with the gene enrichment analysis when it was applied to adverse event enrichment analysis. Unlike the continuous gene expression data, the adverse event data are counts, and a large amount of AEs have a zero count. For example, in the VAERS data, approximately 40\% AEs were never mentioned with the ``FLU4" vaccine, resulting in 40\% AEs with a zero count. The current gene enrichment tests can not handle excessive zeros. Additionally, we encountered 20\% ties in a ratio measure (defined in Methods section) with the count data.  The current GSEA assigns random ranks to the tied statistics, which can lead to inaccurate results. In this work, we extend the current  enrichment tests to appropriately address the above two issues to perform AE enrichment analysis.

\section{Methods}

\paragraph{Data Structure}

For a particular vaccine (denoted as the target vaccine), we create a $2 \times N$ contingency table (see Table \ref{tab:AE&Vax}), with two rows for the target vaccine (Yes/No) and $N$ columns for the AEs reported in the VAERS database during a study period. In this table,  $n_{1i}$ is the number of VAERS reports that mention both the target vaccine and the $i^{th}$ AE in a defined period,  $n_{\cdot i}$ is the total number of reports that mention the $i^{th}$ AE, $n_{1\cdot}$ is the total number of reports that mention the target vaccine, and $n_{\cdot\cdot}$ is the total number of reports in the study period.

\begin{table}[h]
\centering
\footnotesize
\caption{AE count data in a $2 \times N$ table for a target vaccine}
\begin{tabular}{|c|cccc|c|}
\hline
\diagbox{Vaccine}{AE} & $AE_1$ & $AE_2$ & $\cdots$ & $AE_N$ &Total \\
\hline
 Yes & $n_{11}$ & $n_{12}$ & $\cdots$ & $n_{1N}$ & $n_{1\cdot}$ \\
 No &$n_{\cdot 1} - n_{11}$ & $n_{\cdot 2}-n_{12}$ & $\cdots$ & $n_{\cdot N}-n_{1N}$ & $n_{\cdot\cdot}-n_{1\cdot}$\\
\hline
 Total & $n_{\cdot 1}$ & $n_{\cdot 2}$ & $\cdots$ & $n_{\cdot N}$ & $n_{\cdot\cdot}$\\
\hline
\end{tabular}
\label{tab:AE&Vax}
\end{table}

\subsection*{\textbf{AEKS: AE enrichment analysis based on modified K-S statistic}}

In this section, we extend the current GSEA \cite{Subramanian:2005, Mootha:2003} to handle AE data with ties and excessive zero. Poisson distribution has been commonly used to model the $n_{1i}$ \cite{DuMouchel:1999,Szarfman:2002,Huang:2011}, 
\begin{equation}
\label{PoissonModel}
    n_{1i} \sim Poisson(n_{\cdot i}\times \lambda_i),  \mbox{~ for~}  i=1,\cdots,N,
\end{equation}

\noindent where $\lambda_i$ is the reporting ratio (RR) for the $i^{th}$ AE with the target vaccine, with a large value indicating a strong safety signal. RRs are the statistics of interest and we will use their maximum likelihood estimates $\frac{n_{1i}}{n_{\cdot i}}$'s as observed values. Our goal is to determine whether members of a AE group tend to have higher RRs.

\paragraph{Calculate enrichment score for each AE group}
\begin{enumerate}
    \item Rank order the $N$ AEs based on the statistic RR. Assume that there are $J$ distinct RRs ($J \leq N$) and we order the $N$ AEs from the highest to the lowest rank as $L=\{\tilde{AE}_1,\Tilde{AE}_2,\ldots,\Tilde{AE}_J\}$, where $\Tilde{AE}_j=\{AE_{j1},\ldots AE_{jn_j}\}$ is a set of $n_j$ AEs with same RR.
    \item Extend GSEA to handle tied RRs. Given position $i$ in $L$, evaluate the fraction of AEs in group $G$ (``hits") and the fraction of AEs not in $G$ (``misses"). $N_G$ denotes the number of distinct AE terms in group $G$.
         $$P_{hit}(G,i)  = \sum_{j \leq i}\sum_{k\leq n_j }\frac{1}{N_G}I(AE_{jk} \in G)$$
         $$P_{miss}(G,i) = \sum_{j \leq i}\sum_{k\leq n_j }\frac{1}{N-N_G}I(AE_{jk} \notin G) $$
        We then compute a running sum across all $N$ AEs. The K-S statistic for AE group $G$ is defined as 
        $KS(G)=max_{1 \leq i \leq J} (P_{hit}(G,i)-P_{miss}(G,i)),$ which is the maximum value that $P_{hit}$ is above $P_{miss}$.   When many members of $G$ appear at the top of the list, $KS(G)$ is high.
    \item Handle zero counts. The maximum likelihood estimate for $\lambda_i$ is $\frac{n_{1i}}{n_{\cdot i}}.$ Thus, a zero count will produce a zero RR. Let $p_0^G$ denotes the proportion of AEs with  RR equal to $0$ in group $G$, and $p_0^{G^c}$ denotes the proportion not in group $G$. We consider Group $G$  not enriched only if  $p_0^G$ is larger than $p_0^{G^c}.$ 
    \item Combine the statistics in 2 and 3, we propose a composite enrichment score
    \begin{equation}
    \label{new_KS_3}
    \begin{aligned}
     ES(G) &=KS(G) \times I(p_0^G\leq p_0^{G^c}),
    \end{aligned}
    \end{equation}
    
where $ES(G)\in [0,1],$ and $ES(G)$ is zero if the proportion of zero RRs in group $G$ is larger than the proportion of zero RRs in other groups. $ES(G)$ is large if group $G$ has a small proportion of zero RRs than the remaining groups and the non-zero RRs in group $G$ are concentrated at the top of the list $L$.
\end{enumerate}

\paragraph{Estimate statistical significance}
The distribution of $ES(G)$ under the null is not analytically tractable and is obtained using Monte Carlo hypothesis testing\cite{Huang:2011}. Under the null hypothesis, $H_0$: $\lambda_1=\lambda_2=\cdots=\lambda_N=\lambda_0$. Under this hypothesis,  $ n_{1i} \sim Poisson(n_{\cdot i}\times \lambda_0), \mbox{~for~}  i=1,\cdots,N$. Based on the relationship between Poisson and Multinonial distributions, the joint distribution of $(n_{11},\ldots,n_{1N})$, conditioning on $\sum_{i=1}^N n_{1i}= n_{1\cdot}$ and $(n_{\cdot 1},\ldots,n_{\cdot N})$ is
\begin{equation}\label{multinomial}
     (n_{11},n_{12},\ldots,n_{1N})|n_{1\cdot};n_{\cdot 1},\ldots,n_{\cdot N}\sim \mbox{multinomial}\left(n_{1\cdot}, \{r_1,\ldots,r_N\}\right)
\end{equation}

where $r_i= \frac{n_{\cdot i}\lambda_0}{n_{\cdot\cdot} \lambda_0} =\frac{n_{\cdot i}}{n_{\cdot\cdot} }.$

\noindent Given this Multinomial distribution, we generate the AE count data and  compute  $ES^{*}(G)$ using formula (\ref{new_KS_3}). Repeat this process for $M$ times ($M$ is generally large; say 5000) to create a null distribution  of  $ES^{*}(G)$. The $p$-value is the proportion of $ES^{*}(G)$ that are greater than or equal to the observed $ES(G)$.  Finally, we apply the Benjamini-Hochberg procedure \cite{Thissen:2002} converting $p$-values into $q$-values to control the false discovery rate. Moreover, we can also estimate the statistical significance of each individual AE by evaluating the $p$-value of observed RR under the null hypothesis.


\subsection*{\textbf{AEFisher: AE Enrichment test based on modified Fisher's exact test}}

This approach first assesses significance of the association between each AE and the vaccine and then uses a ``cutoff" strategy to classify the AEs into signaled and unsignaled AEs. To test significance of the association, we apply the Fisher's exact test to data in 2 by 2 table (see Table \ref{tab:oneAE&Vax}) and then use the Benjamini-Hochberg procedure to convert $p$-values into $q$-values for controlling the false discovery rate. A signaled AE is defined based on both the strength of the signal and statistical significance, such as $q$-value $<0.1$ and odds ratio (OR) $>1.5$.

\begin{table}[h]
\centering
\caption{A 2 by 2 contingency table for a vaccine-$AE_i$ pair}
\footnotesize
\begin{tabular}{|c|cc|}
\hline
\diagbox{Vaccine}{$AE_i$} & Yes & No \\
\hline
 Yes & $n_{1i}$ & $n_{\cdot i}-n_{1i}$  \\
 No & $n_{\cdot i}-n_{1i}$ & $(n_{\cdot\cdot}-n_{1\cdot})-(n_{\cdot i}-n_{1i})$\\
\hline
\end{tabular}
\label{tab:oneAE&Vax}
\end{table}

To conduct the enrichment analysis for a particular AE group $G$, a conventional approach is to compare proportions of the signaled AEs in group $G$ and not in group $G$. If there are significantly more signaled AEs in group $G$, then group $G$ is enriched. To incorporate the excessive zero RRs in the test, we propose a composite enrichment score as 
$$ES(G) = OR^G\cdot I(p_0^G\leq p_0^{G^c}), $$

\noindent where $OR^G$ is the odds ratio estimating the association between signaled AEs and group $G$. A large $OR^G$ ($OR^G >1$) indicates more signaled AEs in group $G$ than in the remaining groups.  As in the AEKS test, $I(p_0^G\leq p_0^{G^c})$ ensures that an enriched group has the proportion of zeros smaller or equal to remaining groups. 

\paragraph{Estimate statistical significance} We perform a permutation test to assess  significance of the enrichment score for group $G$ by randomly reshuffling the signaled/unsignaled labels. This in spirit is the same as the Fisher's exact test of fixing the row and column margins (here, the  group size and the total number of signaled and unsignaled AEs are fixed), while considering the zero proportions.

\section{Results and Discussion}

\subsection{Simulation studies}

We ran simulation studies to investigate our proposed methods and compared them to existing enrichment tests. To make simulation studies more realistic, data in simulated datasets were made similar to the real dataset. We first created the AE group structure using the AE groups defined on the HLGT level in MedDRA. In each simulated dataset, we set the number of AE groups to be 150 and determined the group size, $N_G,$ by randomly sampling the group size data in MedDRA under the constraint of $N_G\geq 10$. Similarly, the total count of each AE was determined by randomly sampling the AE total count data in VAERS. Then we randomly selected 10\% of the AE groups as enriched and the remaining groups as un-enriched. The proportion of zero AEs per group ($p_0$) is between 10-60\% in VAERS, therefore, we used $p_0$ in this range in simulations. For a non-zero AE count,  we generated it from a Poisson distribution in (\ref{PoissonModel}) with the rate parameter randomly sampled from the estimated $\lambda$'s in VAERS. Specifically, in an enriched group, $p_0$ was sampled uniformly from 0.1 to 0.3 and the rate parameter was constrained to be larger than 0.3. In an un-enriched AE group,  we either set the range of $p_0$ between 0.4 to 0.6 without constraining the rate parameter, or set the rate parameter smaller than 0.4 with $p_0$ in the range of 0.1 to 0.6.

\begin{figure}[h!]
\centerline{
\includegraphics[scale=0.4]{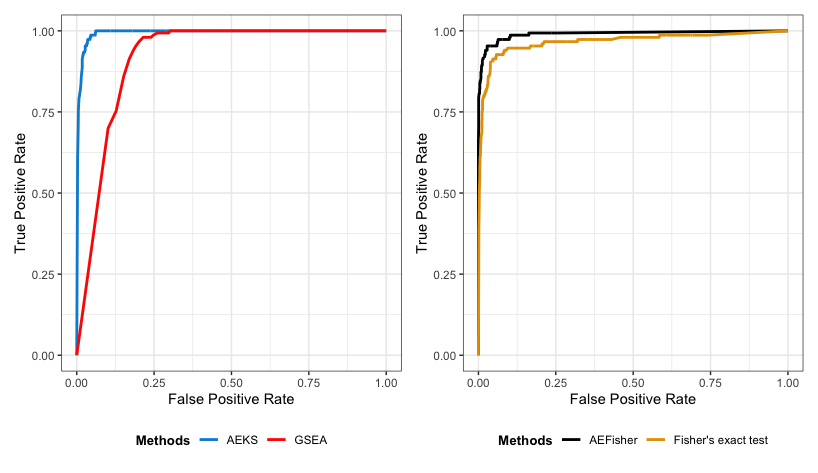}}
\caption{\footnotesize ROC curves using K-S statistics (AEKS vs GSEA) (left) and Fisher's exact tests (AEFisher vs Fisher's exact test) (right). In the Fisher exact test, a signaled AE is defined if $q$-value $<0.1$ and OR $>1.5$. Results were summarized over 10 simulated datasets.}
\label{HLGT_ROC}
\end{figure}

As shown in Figure~\ref{HLGT_ROC}, AEKS and the AEFisher performed significantly better than the GSEA and the conventional Fisher's exact test, respectively. 

\subsection{Application to VAERS datasets}

We applied AEKS and AEFisher to VAERS dataset to study flu and hepatitis vaccines. In both studies, we used the HLGT level of MedDRA to define AE groups. In AEFisher, a signaled AE is defined if the $q$-value $<0.1$ and OR $>1.5$. In both AEKS and AEFisher, an AE group is significantly enriched if $q$-value $<0.1$.  

\subsubsection{Study flu vaccines}
Influenza vaccine is given in large quantities and it prevents millions of illnesses and flu-related doctor’s visits each year. CDC recommends the appropriate vaccine during the flu season. Options include
inactivated influenza vaccine (IIV) (``FLU3" or ``FLU4" in VAERS) or live attenuated influenza vaccine (LAIV) (``FLUN3" or ``FLUN4" in VAERS).  By restricting the  age of the vaccine recipients between 2 and 49, we have 139353 and 21820 reports for IIV and LAIV, respectively. We compared AE profiles with LAIV relative to IIV. 

As shown in Table \ref{tab:FLUN_FLU_res}, AEKS and AEFisher identified the same enriched AE groups: Respiratory tract infections and Upper respiratory tract disorders. Relative to IIV, LAIV is associated with increased risk of respiratory system disorders. Individual AE identified in each group include Rhinitis, Nasal congestion, Sinus disorder, which have been reported before \cite{Baxtera:2017,Haber:2015,Rob:2016}, New signals, such as Epistaxis, is clinically interesting, and it might be true  signals that need to be validated in large healthcare databases.

\begin{table}[h!]
\caption{The enriched AE groups and significant AEs using AEKS and AEFisher to study LAIV relative to IIV.}
\centering
\footnotesize
\begin{tabular}{cll}
\hline
\multicolumn{1}{l}{Methods} & Significant groups & Significant AEs \\ \hline
\multirow{2}{*}{\begin{tabular}[c]{@{}c@{}} \\ \\ \\ AEKS \\ \& \\ AEFisher\end{tabular}} & Respiratory tract infections & \begin{tabular}[c]{@{}l@{}}Croup infectious\\ Influenza\\ Nasopharyngitis\\ Pneumonia\\ Rhinitis\\ Sinusitis\\ Upper respiratory tract infection\end{tabular} \\ \cline{2-3} 
 & \begin{tabular}[c]{@{}l@{}}Upper respiratory tract\\ disorders (excl infections)\end{tabular} & \begin{tabular}[c]{@{}l@{}}Anosmia\\ Epistaxis\\ Nasal congestion\\ Nasal oedema\\ Seasonal allergy\\ Sinus disorder\\ Stridor\\ Tonsillar hypertrophy\end{tabular} \\ \hline
\end{tabular}
\label{tab:FLUN_FLU_res}
\end{table}

\subsubsection{Study hepatitis A and B combination vaccines}

In this study, we are interested in identifying safety problems that are likely due to interactions of two vaccines when they are administered to an individual at the same time. Specifically, we compared  AE profiles induced by the hepatitis A and B combination vaccine (``Twinrix" in VAERS) to monovalent hepatitis A and B vaccines (``Havrix" for hepatitis A and ``Engerix-B" for hepatitis B in VAERS). We selected vaccine reports from 2002 to 2018. There are 53415, 33087, 10356 reports with Havrix, Engerix-B, and Twinrix, respectively.  In this study, AEKS and AEFisher identified different AE groups. As shown in Table \ref{tab:HEPAB}, AEKS identified Peripheral neuropathies, while AEFisher identified Musculoskeletal and connectivetissue disorders NEC. Peripheral neuropathies was also mentioned in  \cite{Xie:2016} as an important AE group associated with the combination hepatitis vaccine.  

\begin{table}[h!]
\caption{The enriched AE groups and significant AEs using AEKS and AEFisher to study hepatitis A and B combination relative to monovalent Hepatitis A and B vaccines}
\centering
\footnotesize
\begin{tabular}{lll}
\hline
Methods & Significant AE group & Significant AEs \\ \hline
AEKS & Peripheral neuropathies & \begin{tabular}[c]{@{}l@{}}Guillain-Barre syndrome \\ Miller Fisher syndrome \\ Neuropathy peripheral\end{tabular} \\ \hline
AEFisher & \begin{tabular}[c]{@{}l@{}}Musculoskeletal and connective\\ tissue disorders NEC\end{tabular} & \begin{tabular}[c]{@{}l@{}}Back pain \\ Mobility decreased \\ Musculoskeletal disorder \\ Musculoskeletal pain  \\ Neck pain \\ Pain in extremity\end{tabular} \\ \hline
\end{tabular}
\label{tab:HEPAB}
\end{table}

\section{Conclusions}
In this article, we develop new methods specifically for adverse event enrichment analysis. We extend the existing gene enrichment tests by incorporating the special features of the AE count data.  We have demonstrated the advantage of our methods over the existing methods in simulation studies. By applying the proposed methods to VAERS data, we found that relative to IIV, LAIV is associated with increased risk of respiratory system disorders, and the hepatitis A and B combination vaccine, relative to monovalent Hepatitis A or B vaccines, may be associated with Peripheral neuropathies or Musculoskeletal and connective problems.

\section{Acknowledgements}
Thanks to Kirsten Herold for proofreading the article and thanks to Michael Kleinsasser for compiling the R package.

\bibliographystyle{unsrt}  
\bibliography{EnrichAE}

\begin{thebibliography}{10}

\bibitem{Varricchio:2004}
F.~Varricchio, J.~Iskander, F.~DeStefano, R.~Ball, R.~Pless, M.~M. Braun, and
  R.~T. Chen.
\newblock Understanding vaccine safety information from the vaccine adverse
  event reporting system.
\newblock {\em Pediatr Infect Dis J.}, 23:287–--294, 2004.

\bibitem{Shimabukuro:2015}
Tom~T Shimabukuro, Michael Nguyen, David Martin, and Frank DeStefano.
\newblock Safety monitoring in the vaccine adverse event reporting system
  (vaers).
\newblock {\em Vaccine}, 33(36):4398--4405, 2015.

\bibitem{DuMouchel:1999}
W.~DuMouchel.
\newblock Bayesian data mining in large frequency tables, with an application
  to the {FDA} spontaneous reporting system.
\newblock {\em The American Statistician}, 53:177–190, 1999.

\bibitem{Evans:2001}
S.~J. Evans, P.~C. Waller, and S.~Davis.
\newblock Use of proportional reporting ratios ({PRR}s) for signal generation
  from spontaneous adverse drug reaction reports.
\newblock {\em Pharmacoepidemiology and Drug Safety}, 10:483--486, 2001.

\bibitem{van:2002}
E.~P. van Puijenbroek, A.~Bate, H.~G. Leufkens, M.~Lindquist, R.~Orre R, and
  A.~C. Egberts.
\newblock A comparison of measures of disproportionality for signal detection
  in spontaneous reporting systems for adverse drug reactions.
\newblock {\em Pharmacoepidemiology and Drug Safety}, 11:3--10, 2002.

\bibitem{Bate:1998}
A.~Bate, M.~Lindquist, I.~R. Edwards, S.~Olsson, R.~Orre, A.~Lansner, and
  R.~M.~De Freitas.
\newblock A {B}ayesian neural network method for adverse drug reaction signal
  generation.
\newblock {\em European Journal of Clinical Pharmacology}, 54:315–321, 1998.

\bibitem{Orre:2000}
R.~Orre~A. Lansner, A.~Bate, I.~R.~Edwards M.~Lindquist, S.~Olsson, R.~Orre,
  A.~Lansner, and R.~M.~De Freitas.
\newblock Bayesian neural networks with confidence estimations applied to data
  mining.
\newblock {\em Comput Stat Data Anal}, 34:473–493, 2000.

\bibitem{Noren:2006}
G.~N. N{\'o}ren, A.~Bate, R.~Orre, and I.~R. Edwards.
\newblock Extending the methods used to screen the {WHO} drug safety database
  towards analysis of complex associations and improved accuracy for rare
  events.
\newblock {\em BCPNN}, 25:3740–3757, 2006.

\bibitem{DuMouchel:2001}
W.~DuMouchel and D.~Pregibon.
\newblock Empirical bayes screening for multi-item associations.
\newblock {\em Proceedings of the Seventh ACM SIGKDD International Conference
  on Knowledge Discovery and Data Mining, San Francisco,CA}, page 67–76,
  2001.

\bibitem{Szarfman:2002}
A.~Szarfman, S.~G. Machado, and R.~T. O’Neill.
\newblock Use of screening algorithms and computer systems to efficiently
  signal higher-than-expected combinations of drugs and events in the us
  {FDA}’s spontaneous reports database.
\newblock {\em PLoS ONE}, 25:381–392, 2002.

\bibitem{Kulldorff:2011}
M.~Kulldorff, R.~L. Davis, M.~Kolczak, E.~Lewis, T.~Lieu, and R.~Platt.
\newblock A maximized sequential probability ratio test for drug and vaccine
  safety surveillance.
\newblock {\em Sequential Analysis}, 30:58–78, 2011.

\bibitem{Davis:2005}
R.~L. Davis, M.~Kolczak, E.~Lewis, J.~Nordin, M.~Goodman, D.~K. Shay, R.~Platt,
  S.~Black, H.~Shinefield, and R.~T Chen.
\newblock Active surveillance of vaccine safety: a system to detect early signs
  of adverse events.
\newblock {\em Epidemiology}, 16:336–341, 2005.

\bibitem{Li:2009}
L.~Li and M.~Kulldorff.
\newblock A conditional maximized sequential probability ratio test for
  pharmacovigilance.
\newblock {\em Statistics in Medicine}, 29:284–295, 2009.

\bibitem{Li:2014}
Rongxia Li, Brock Stewart, Eric Weintraub, , and Michael~M. McNeil.
\newblock Continuous sequential boundaries for vaccine safety surveillance.
\newblock {\em Statistics in Medicine}, 33:3387--3397, 2014.

\bibitem{Kulldorff:2013}
M~Kulldorff, I.~Dashevsky, T.~R. Avery, K.~A. Chan, R.~L. Davis, D.~Graham,
  R.~Platt, S.~E. Andrade, D.~Boudreau, M.~J. Gunter, L.~J. Herrinton,
  P.~Pawloski, M.~A. Raebel, D.~Roblin, and J.~S. Brown.
\newblock Drug safety data mining with a tree-based scan statistic.
\newblock {\em Pharmacoepidemiology and Drug Safety}, 22:517--523, 2013.

\bibitem{Mozzicato:2009}
P.~Mozzicato.
\newblock Meddra: an overview of the medical dictionary for regulatory
  activities.
\newblock {\em Pharm Med}, 23:65–--75, 2009.

\bibitem{Zhang:2013}
Y.~Zhang, C.~Tao, Y.~He, P.~Kanjamala, and H.~Liu H.
\newblock Network-based analysis of vaccine-related associations reveals
  consistent knowledge with the vaccine ontology.
\newblock {\em Journal of Biomedical Semantics}, 4:33, 2013.

\bibitem{Marcos:2013}
Erica Marcos, Bin Zhao, and Yongqun He.
\newblock The ontology of vaccine adverse events ({OVAE}) and its usage in
  representing and analyzing adverse events associated with us-licensed human
  vaccines.
\newblock {\em Journal of Biomedical Semantics}, 4:40, 2013.

\bibitem{Sarntivijai:2012}
S.~Sarntivijai, Z.~Xiang, K.~A. Shedden, H.~Markel, G.~S. Omenn, B.~D. Athey,
  and Y.~He.
\newblock Ontology-based combinatorial comparative analysis of adverse events
  associated with killed and live influenza vaccines.
\newblock {\em PLoS ONE}, 7:e49941, 2012.

\bibitem{Guo:2016}
A.~Guo, R.~Racz, J.~Hur, Y.~Lin, Z.~Xiang, L.~Zhao, J.~Rinder, G.~Jiang,
  Q.~Zhu, and Y.~He.
\newblock Ontology-based collection, representation and analysis of
  drug-associated neuropathy adverse events.
\newblock {\em J Biomed Semantics.}, 7:29, 2016.

\bibitem{Zeeberg:2003}
Barry~R Zeeberg, Weimin Feng, Geoffrey Wang, May~D Wang, Anthony~T Fojo, Margot
  Sunshine, Sudarshan Narasimhan, David~W Kane, William~C Reinhold, Samir
  Lababidi, et~al.
\newblock Gominer: a resource for biological interpretation of genomic and
  proteomic data.
\newblock {\em Genome biology}, 4(4):R28, 2003.

\bibitem{Subramanian:2005}
Aravind Subramanian, Pablo Tamayo, Vamsi~K Mootha, Sayan Mukherjee, Benjamin~L
  Ebert, Michael~A Gillette, Amanda Paulovich, Scott~L Pomeroy, Todd~R Golub,
  Eric~S Lander, et~al.
\newblock Gene set enrichment analysis: a knowledge-based approach for
  interpreting genome-wide expression profiles.
\newblock {\em Proceedings of the National Academy of Sciences},
  102(43):15545--15550, 2005.

\bibitem{Mootha:2003}
Vamsi~K Mootha, Cecilia~M Lindgren, Karl-Fredrik Eriksson, Aravind Subramanian,
  Smita Sihag, Joseph Lehar, Pere Puigserver, Emma Carlsson, Martin
  Ridderstr{\aa}le, Esa Laurila, et~al.
\newblock Pgc-1$\alpha$-responsive genes involved in oxidative phosphorylation
  are coordinately downregulated in human diabetes.
\newblock {\em Nature genetics}, 34(3):267--273, 2003.

\bibitem{Huang:2011}
Lan Huang, Jyoti Zalkikar, and Ram~C Tiwari.
\newblock A likelihood ratio test based method for signal detection with
  application to fda’s drug safety data.
\newblock {\em Journal of the American Statistical Association},
  106(496):1230--1241, 2011.

\bibitem{Thissen:2002}
David Thissen, Lynne Steinberg, and Daniel Kuang.
\newblock Quick and easy implementation of the benjamini-hochberg procedure for
  controlling the false positive rate in multiple comparisons.
\newblock {\em Journal of educational and behavioral statistics}, 27(1):77--83,
  2002.

\bibitem{Baxtera:2017}
Roger Baxtera, Abigail Eatona, John Hansena, Laurie Aukesa, Herve Caspardb, and
  Christopher~S. Ambroseba.
\newblock Safety of quadrivalent live attenuated influenza vaccinein subjects
  aged 2–49 years.
\newblock {\em Vaccine}, 35:1254--1258, 2017.

\bibitem{Haber:2015}
Penina Haber, Pedro~L. Moro, Maria Cano, Paige Lewis, Brock Stewart, and Tom
  T.Shimabukuro.
\newblock Post-licensure surveillance of quadrivalent live attenuated influenza
  vaccine united states,vaccine adverse event reporting system({VAERS}), july
  2013–june 2014.
\newblock {\em Vaccine}, 33:1987--1992, 2015.

\bibitem{Rob:2016}
Rob Lambkin-Williams, Colin Gelder, Richard Broughton, Corey~P. Mallett,
  Anthony~S. Gilbert, Alex Mann, David He, John~S. Oxford, and David Burt.
\newblock An intranasal proteosome-adjuvanted trivalent influenza vaccine is
  safe, immunogenic and efficacious in the human viral influenza challenge
  model. serum {IgG} and mucosal {IgA} are important correlates of protection
  against illness associated with infection.
\newblock {\em PLoS One}, 11:e0163089, 2016.

\bibitem{Xie:2016}
J.~Xie, L.~Zhao, S.~Zhou, and Y.~He.
\newblock Statistical and ontological analysis of adverse events associated
  with monovalent and combination vaccines against hepatitis {A} and {B}
  diseases.
\newblock {\em Scientific Report}, 6:3418, 2016.

\end{thebibliography}

\end{document}